\documentclass[%
 reprint,
superscriptaddress,
 amsmath,amssymb,
 aps,
pra,
]{revtex4-1}

\usepackage{graphicx}
\usepackage{dcolumn}
\usepackage{bm}

\begin{document}


\title{Degenerate four-wave-mixing in a silica hollow bottle-like microresonator}

\author{Yong Yang}
 \affiliation{Light-Matter Interactions Unit, Okinawa Institute of Science and Technology Graduate University, Onna, Okinawa 904-0495, Japan}
  \affiliation{National Engineering Laboratory for Fiber Optics Sensing Technology, Wuhan University of Technology, Wuhan, 430070, China
}%
   \email{yong.yang@oist.jp}
\author{Yuta Ooka}%
\affiliation{%
Light-Matter Interactions Unit, Okinawa Institute of Science and Technology Graduate University, Onna, Okinawa 904-0495, Japan}%
\affiliation{Dept. of Electronics and Electrical Engineering, Faculty of Science and Technology, Keio University, 3-14-1, Hiyoshi, Kohoku-ku, Yokohama 223-8522, Japan}
\author{Ruth Thompson}
\affiliation{Light-Matter Interactions Unit, Okinawa Institute of Science and Technology Graduate University, Onna, Okinawa 904-0495, Japan}%
\author{Jonathan Ward}
\affiliation{Light-Matter Interactions Unit, Okinawa Institute of Science and Technology Graduate University, Onna, Okinawa 904-0495, Japan}%
\author{S\'ile Nic Chormaic}
\affiliation{Light-Matter Interactions Unit, Okinawa Institute of Science and Technology Graduate University, Onna, Okinawa 904-0495, Japan}

\date{\today}

\begin{abstract}
A hollow bottle-like microresonator (BLMR) with ultra-high quality factor is fabricated from a microcapillary with nearly parabolic profile. At 1.55 $\mu m$ pumping, degenerate four-wave mixing can be observed for a BLMR of diameter 102 $\mu$m. The parabolic profile of the BLMR guarantees a nearly zero waveguide dispersion, which is theoretically discussed in detail. From the simulation, at 1.55 $\mu$m wavelength in such a BLMR, the fundamental bottle mode is in the anomalous dispersion regime, whilst the ordinary whispering gallery mode (WGM) confined at the center of the BLMR is in the normal dispersion regime. Experimentally, no degenerate FWM is observed for the WGM selected by positioning the coupling tapered fiber in the same BLMR. Furthermore, dispersion tuning is briefly discussed. As the work predicted, the BLMR shows promise for the implementation of sparsely distributed, widely spanned frequency combs at the telecommunication wavelength.
\end{abstract}

\maketitle


\section{Introduction}
Whispering gallery mode microresonators (WGRs) are widely used as a basic element for many applications in various fields such as cavity quantum electrodynamics\cite{Aoki2006}, optomechanics\cite{Kippenberg2008} and sensing\cite{Vollmer2008}. The photons in the WGRs are highly confined by total internal reflection, so that the light intensity in the WGRs can be very high. Therefore many non-linear effects such as Raman\cite{Spillane2002}, Brillouin scattering\cite{Bahl2011} and Kerr effects\cite{Pollinger2010} can be implemented. Among the different non-linear effects, four-wave-mixing (FWM) is very important in generating frequency combs\cite{Kippenberg2011} and cavity solitons\cite{Herr2013}. FWM is a parametric process where two pumping photons create a signal-idler pair, which should satisfy conservation of energy and momentum\cite{Sharping2002}. In an ideal cavity, the cavity modes should be equidistant, such that if we pump the cavity at one mode resonance, the adjacent modes symmetric to this mode automatically satisfy the requirements (i.e. they are phase matched). Such a process is called degenerate FWM. However, in reality cavity modes are not equidistant due to dispersion. Dispersion originates from two sources; material dispersion, which can be quantized by the parameter of group velocity dispersion (GVD) if we only consider up to second order dispersion in the material\cite{DelHaye2007}, and waveguide dispersion\cite{Ilchenko2003}, introduced by the non-equidistant property of the cavity.

In WGRs, in order to obtain the phase matching condition under dispersion for degenerate FWM, optical Kerr non-linearity in the cavity is adapted\cite{Kippenberg2004, Agha2007}. Taking silica WGRs as an example, the material non-linear index coefficient is positive such that the total dispersion needs to be anomalous. The zero dispersion wavelength (ZDW) of silica is approximately 1.33 $\mu$m; thus, at the telecommunication wavelength (1.55 $\mu$m) it is anomalous\cite{DelHaye2007}. In a silica microsphere WGR, the waveguide dispersion will shift the ZDW to longer wavelengths. In order to maintain anomalous dispersion, the size of the microsphere should be chosen carefully. It was demonstrated that in a silica microsphere of diameter no less than 150 $\mu$m, the total dispersion is still in the anomalous regime and therefore FWM can be observed in such microspheres\cite{Agha2007}. Recently, it has been found that microubble WGRs may be a good candidate for compensating for waveguide dispersion. With a thin wall, the cavity mode is highly confined, such that the waveguide dispersion can be reduced. In a microbubble with a wall thickness of 3-4 $\mu$m, FWM was observed even for diameters at 136 $\mu$m\cite{Li2013}.

Waveguide dispersion in a microsphere or microbubble originates from the non-equidistance of the mode distribution for different longitudinal modes. From the Mie scattering model of a spherical boundary, the eigenvalues of the equations give a non-linear relationship to the longitudinal mode numbers\cite{Gorodetsky2006}. To get equally distributed eigenvalues, it was proposed that bottle-like mode resonators (BLMR) could be used\cite{Sumetsky2004}. Unlike conventional WGRs, where light is confined in the equatorial region, in such a BLMR light propagates not only in the equatorial plane but also along the axial direction of the cavity, spiraling between two caustics. Q factors as high as $10^8$ were achieved in BLMRs experimentally\cite{Pollinger2009}, allowing their use in non-linear optics\cite{Ooka2015}. If the lateral profile of the BLMR is parabolic, theoretically the equation describing the axial electromagnetic fields of the bottle like modes is an analogy of a harmonic oscillator; thus, the eigenvalue is ideally linear to the index of the mode numbers.

In this letter, we analyzed the waveguide dispersion of a BLMR and compared to the WGM without axial propagation. By using the two CO$_2$ beam heating technique, a hollow BLMR was fabricated. It is shown that degenerated FWM at 1.55 $\mu$m wavelength exists for a BLMR smaller than that for a microsphere or microbubble if the bottle modes are excited by selective coupling using a tapered fiber.

\begin{figure}[h]
\centering\includegraphics[width=.9\columnwidth]{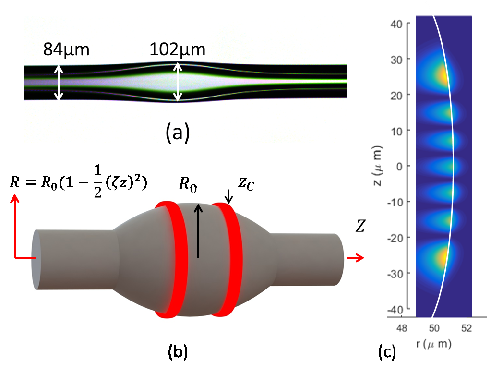}
\caption{(a) Microscope image of the BLMR for the experiment. (b) A scheme of a bottle mode ($m=1$) in a cylindrical coordinate. $R_0$ is the radius at the center. $z_C$ is the coordinate for the caustic.(c) is a $m=6$ mode pattern in a $R_0=51 \mu m$ BLMR.}
\label{fig:bottle}
\end{figure}
\section{Dispersion in BLMR}
The BLMR can be treated as a cylindrical optical cavity with a lateral deformation illustrated as Fig.\ref{fig:bottle}(b). The profile along the z-axis is a parabolic curve with $R(z)=R_0(1-\frac{1}{2}(\zeta z)^2)$. Here, $R_0$ is the maximum radius of the BLMR at $z=0$, and $\zeta$ is the curvature of the profile. When light is coupled at the caustic $z_c$, the bottle mode is excited. The eigenfrequency $\nu_{l,m}$ of the bottle modes can be expressed as follows\cite{Sumetsky2004, Pollinger2009}:
\begin{equation}
\nu_{l,m}=\frac{c}{2\pi n}\sqrt{\frac{l^2}{R_0^2}+\frac{(m+\frac{1}{2})2\zeta l}{R_0}}.
\end{equation}
Here, $l,m$ are longitudinal and axial mode numbers respectively, where $m+1$ is the number of round trips between two caustics. $c$ is the speed of light in vacuum and $n$ is the refractive index of the resonator material.

Note that in the practice, $\zeta R_0$ is very small and $m+\frac{1}{2}<l$ such that the waveguide dispersion of the bottle mode, which is defined here as the variation of the free spectrum range (FSR), is:
\begin{equation}
\Delta \nu_{FSR}\approx-\frac{c[(q+\frac{1}{2})\zeta R_0]^2}{2\pi nR_0}l^{-3}.
\label{BLMDispersion}
\end{equation}

If light is coupled at the equatorial plane $z=0$, the modes excited in the BLMR is the conventional WGM as for a microsphere. The TE mode distribution has the following form\cite{Gorodetsky2006}:
\begin{equation}
\nu_{q,l}\approx\frac{c}{2\pi R_0n}[l-\alpha_q(\frac{l}{2})^{\frac{1}{3}}+\frac{3\alpha_q^2}{20}(\frac{l}{2})^{-\frac{1}{3}}].
\label{WGMmode}
\end{equation}
Such a mode is indexed by the radial and longitudinal mode numbers as $q,l$. We only consider the first radial order such that $\alpha_q=2.45$ is the first root of the Airy equation. The dispersion induced FSR variation can be calculated as\cite{Kippenberg2004}:
\begin{equation}
\Delta\nu_{FSR}\approx-\frac{0.41c}{2\pi nR_0}l^{-5/3}.
\label{WGMDispersion}
\end{equation}

\begin{figure}[h]
\centering\includegraphics[width=.8\columnwidth]{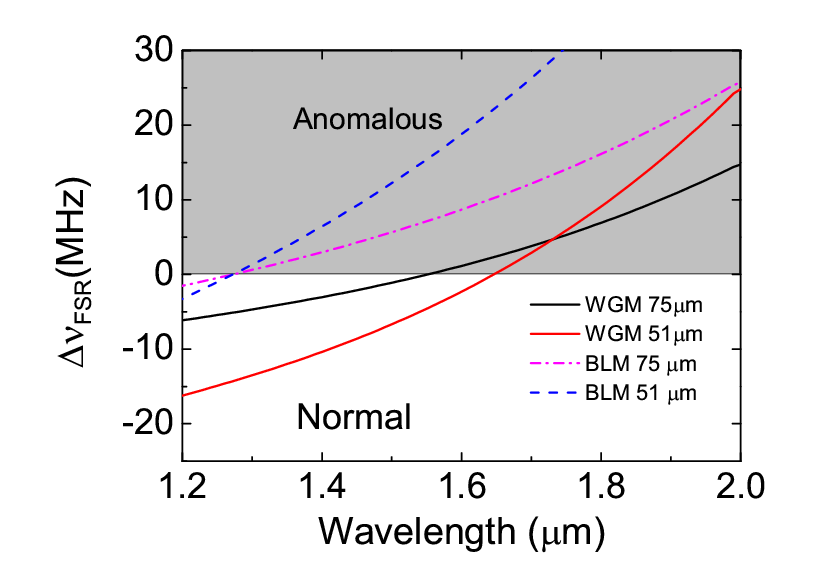}
\caption{The dispersion of a BLMR in terms of FSR variation for different working wavelengths. The solid curves are for WGMs and the dashed curves represent the bottle modes. Two different radii are chosen as illustrated in the legends.}
\label{fig:totaldispersion}
\end{figure}

The total dispersion of the cavity mode includes both material dispersion and waveguide dispersion. The variation of FSR due to material dispersion is $\Delta\nu^0_{FSR}\approx c^2\lambda^2/4\pi^2n^3R^2\cdot GVD$. $GVD=-(\lambda/c)(\partial^2n/\partial\lambda^2)$\cite{DelHaye2007}. Using Eq.\ref{BLMDispersion}, \ref{WGMDispersion} the total variation can be calculated. For example, in Fig. \ref{fig:totaldispersion}, the dispersion of the bottle mode and WGM for BLMRs of radius 51 $\mu m$ and 75 $\mu$m is illustrated. It can be seen that for the bottle mode, waveguide dispersion does not shift the ZDW significantly, whilst WGM dispersion shifts the ZDW depending on the radius of the BLMR. We then plotted the FSR variation for different sizes of BLMR at 1.55 $\mu$m wavelength, as illustrated in Fig.\ref{fig:totaldispersion}. At a radius of 75 $\mu$m, the WGM dispersion induced variation exceeds zero, as shown in \cite{Li2013}. For a radius of 51 $\mu$m the FSR variation is $-5 MHz$, which means that dispersion is in the normal regime. Degenerate FWM should not be observed due to the invalidity of the phase matching condition, which has been experimentally demonstrated in \cite{Agha2007, Li2013}. However, for the $m=1$ bottle mode, the variation is always positive for any radius, such that the phase matching can be satisfied. Degenerate FWM should therefore be observed in a BLMR at a radius that in conventional WGMs is forbidden. This has been experimentally verified as follows.

\begin{figure}[h]
\centering\includegraphics[width=0.8\columnwidth]{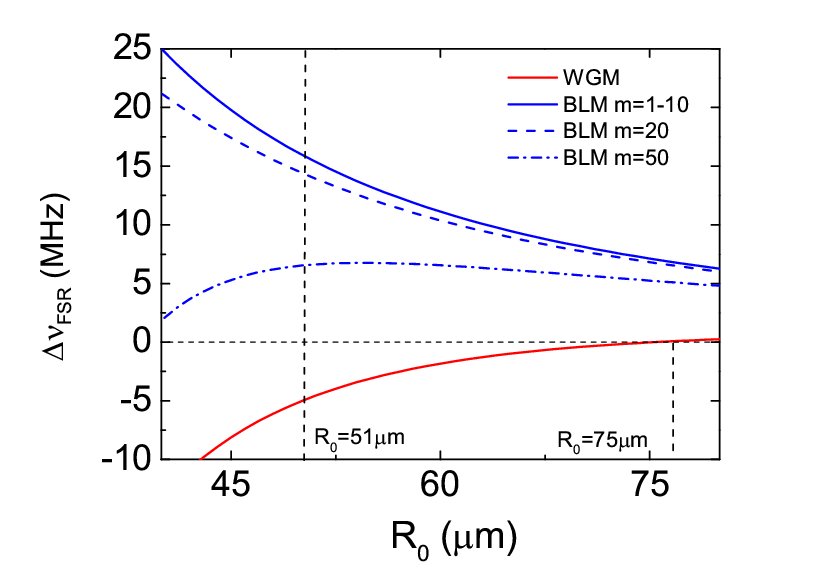}
\caption{The dispersion by variation of FSR of BLMR with different sizes at 1.55 $\mu$m wavelength. Red curves correspond to conventional WGMs and blue curves represent the bottle modes of $m=1-10, 20, 50$ separately.}
\label{fig:dispersionsize}
\end{figure}
\section{Experiment}
BLMRs can be fabricated from controllable tapering using an optical fiber\cite{Pollinger2009, Ward2006}. Here, we fabricated the BLMR from a silica microcapillary using the two-beam CO$_2$ laser technique\cite{Ooka2015}. A microcapillary with inner and outer diameters of 100 $\mu$m and 350 $\mu$m respectively was tapered to an outer diameter of about 90 $\mu$m. The tapered microcapillary was then placed in the foci of two opposing CO$_2$ beams, whilst connected to a nitrogen gas cylinder with a pressure of approximately 3 bars. The opposing beams evenly heat and soften the microcapillary, and the resulting swelling is monitored with a CCD camera. For obtaining a BLMR, a small swelling of about 9 $\mu$m in radius is sufficient (c.f. Fig.\ref{fig:bottle}(a)). With this method, a near-parabolic BLMR profile can be obtained with a curvature of about $\zeta=0.005\mu m^{-1}$. The wall thickness of the BLMR was estimated to be 15 $\mu$m from under a microscope, meaning that the low radial order modes are mostly confined on the outer surface such that a hollow BLMR can be equivalent to a solid one.

\begin{figure}[h]
\centering\includegraphics[width=\columnwidth]{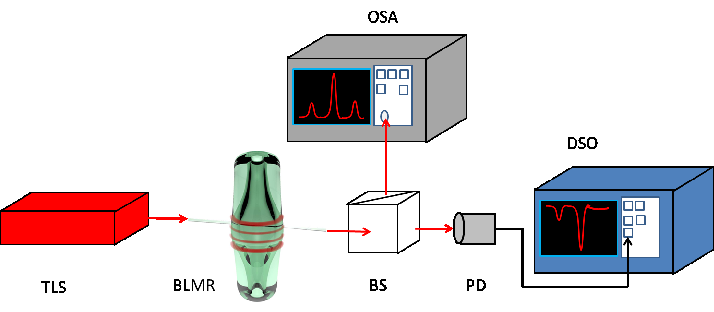}
\caption{Experimental setup for observing FWM in a BLMR. TLS: tunable laser source at 1.55 $\mu$m. BS: beam splitter. OSA: optical spectrum analyzer. PD: photo detector. DSO: digital oscilloscope.}
\label{fig:setup}
\end{figure}

Fig.\ref{fig:setup} shows a schematic for the experimental setup. A tunable laser source at 1.55 $\mu$m was launched into a tapered optical fiber with a waist diameter of 1.3-1.6 $\mu$m made using the heat-and-pull technique\cite{Ward2006}. The BLMR was positioned on a 3D nano-positioner such that it can be evanescently coupled to the tapered fiber. The image of the coupling position was monitored by using a microscope with a high resolution CCD camera. The light coupled from the BLMR was received by a photo detector (PD) and an optical spectrum analyzer (OSA) via a beam splitter. The resulting signal from the PD was displayed on an oscilloscope for finding the optical modes. The tunable laser was set at 1550.15 nm and continuously scanned a 35 GHz range. High Q modes were picked and continuously scanned across. The signal is observed on the OSA as the frequencies are scanned.

\begin{figure}
\includegraphics[width=0.8\columnwidth]{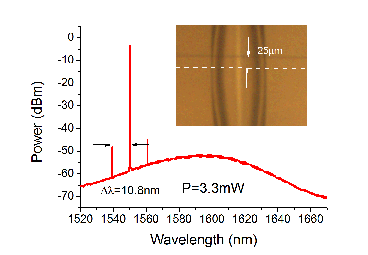}
\caption{The Four-wave-mixing in the hollow BLMR. The pump power is 3.3 mW. The inset shows the position of the taper from the center of the BLMR (dash line) when producing the FWM.}
\label{fig:FWM}
\end{figure}

To excite the bottle mode, the tapered fiber was positioned at approximately 25 $\mu$m from the center, which is the estimated caustic for the $m=6$ bottle mode (see Fig.\ref{fig:bottle}(c)). This position can be measured from the image of the CCD camera (see Fig.\ref{fig:FWM} inset). It can be seen that two sideband peaks of equal height arise alongside the pumping signal in the center. These three wavelengths form the degenerate FWM. The pumping wavelength is at 1550.17 nm and the signal peak is at 1560.95 nm with the idler at 1539.37 nm. The two peaks are equally separated from the pump with an interval of 10.8 nm, in agreement with twice the calculated FSR of the BLMR (5.3 nm). Note that for practical systems, other non-linearities such as Raman scattering also exist. In particular, the Raman process can generate hyper-parametric oscillations even when the dispersion is normal\cite{Agha2007, Farnesi2015}. In order to avoid this, we carefully chose the pump power to be 3.3 mW to 4.2 mW, which is above the FWM threshold but still below the Raman threshold for the bottle mode. To demonstrate this, the OSA was spanned to 1670 nm, where no Raman peaks were found but the FWM.
\begin{figure}[t]
\centering\includegraphics[width=.8\columnwidth]{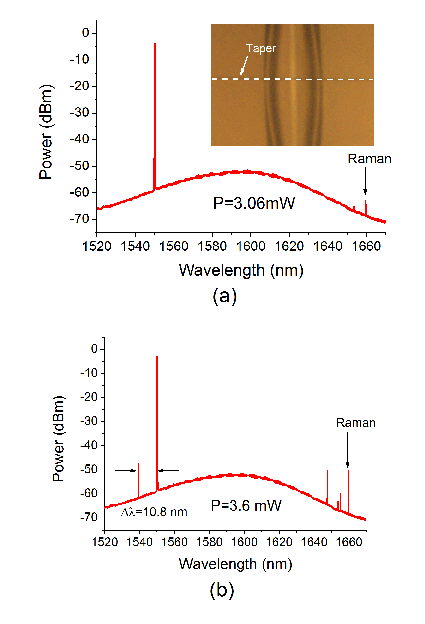}
\caption{Output spectrum when the taper is positioned at the center of the BLMR to excite the conventional WGM ((a), inset). (a) The Raman oscillation in the WGM in the microbottle near the Raman lasing threshold (pump power of 3.06 mW). (b) Non-degenerate FWM assisted by Raman oscillation when the pump power is high (3.6 mW).}
\label{fig:Raman}
\end{figure}

For comparison, the tapered fiber was moved to the center of the BLMR (shown in the inset of Fig.\ref{fig:Raman}), such that the excited mode is the conventional WGM. The same measurements were performed with an input power varying from 1.8 mW to 3.5 mW. From Fig.\ref{fig:Raman}(a), Raman oscillation starts from a pump power of 3.06 mW but no side bands arise. At 3.6 mW, side bands appear; however, these are not symmetric to the pump. Moreover, the long range spectrum shows that the two side bands are separately adjacent to the pump and Raman peak with equal distances of 10.8 nm. This is a non-degenerate FWM assisted by the Raman process\cite{Spillane2002,Min2005}. It is worth noting that Raman oscillation occurs at a lower threshold for WGM than the bottle mode. This is due to a much larger mode volume even for the $m=1$ bottle mode than for the WGM\cite{Ooka2015}. Nevertheless, the degenerate FWM does not exist in the WGM in the same BLMR, which supports the hypothesis that the bottle mode is in the anomalous dispersion regime despite its small size.

For this work, the radius of the BLMR appears to be a good choice. It is in the anomalous regime, but not too deep ($+$15 MHZ for the FSR variation), and with the more than $5\times 10^7$ Q-factor, Kerr non-linearity can just compensate for anomalous dispersion below the Raman threshold. In principle, degenerate FWM exists in arbitrarily small BLMRs as dispersion is mainly dominated by the material at 1.55 $\mu$m. However, it gets into even deeper anomalous dispersion by doing so. This will increase difficulties in phase matching, which requires a higher Q factor or pumping power. The Raman threshold also decreases for smaller BLMRs such that Raman assisted FWM may occur before degenerate FWM.

From Eq.\ref{BLMDispersion}, it can be shown that the waveguide dispersion of the bottle mode is related to the curvature of the profile $\zeta$ and the axial mode number $m$. This means that the system dispersion is manageable. In microbubble resonators, this can be achieved by changing the wall thickness or filling it with a different dispersive material\cite{Li2013}, but both of these methods are at the expense of sacrificing the optical Q-factor. Here, without introducing any other medium and just by selecting a higher axial mode excitation, the dispersion can be reduced (see Fig.\ref{fig:dispersionsize}, while the Q-factor can still be as high as $10^8$ for those modes\cite{Pollinger2009}. We have tried to excite higher order axial modes in our BLMR by moving the tapered fiber even further from the center. High Q modes can still be found, but no degenerate FWM is observed. This is due to deviations of the fabricated profile from the ideal parabolic shape. Since the diameter of the focused CO$_2$ beams are a few tens of $\mu$m, only the central part of the microbottle is well controlled. Therefore, the resulting dispersion may not be in the anomalous regime when coupling far from the center. We believe that by improving control of the fabrication process, higher axial order modes can be used. A new technique called SNAP, where the profile of the cavity can be controlled to sub-Angstrom precision\cite{Sumetsky2012}, could be used for managing dispersion in such devices\cite{Sumetsky2013}.
\section{Conclusion}
In conclusion, we have implemented degenerate FWM in a hollow, parabolic BLMR. The waveguide dispersion of the bottle mode is negligible, and thus the ZDW is determined by the material. Degenerate FWM can be excited for a BLMR of radius 51 $\mu$m, which is unobtainable in the conventional WGM as it is in the normal dispersion regime. These predictions were then verified experimentally. The bottle mode can be in the deep anomalous dispersion regime for very small sizes, such that it could be used for miniaturized or sparsely spaced frequency combs in the future. The hollow structure may also be used for fine tuning the frequency combs generated via aerostatic pressure tuning.
\section*{Acknowledgment}
This work is funded by the Okinawa Institute of Science and Technology Graduate University. The authors thanks Dr. Yongping Zhang and Mr. Ramgopal Madugani for discussions. Mr. L. H. Vu and Mr. K. Vikraman helped us in setting up the experimental system.

\begin{thebibliography}{23}%
\makeatletter
\providecommand \@ifxundefined [1]{%
 \@ifx{#1\undefined}
}%
\providecommand \@ifnum [1]{%
 \ifnum #1\expandafter \@firstoftwo
 \else \expandafter \@secondoftwo
 \fi
}%
\providecommand \@ifx [1]{%
 \ifx #1\expandafter \@firstoftwo
 \else \expandafter \@secondoftwo
 \fi
}%
\providecommand \natexlab [1]{#1}%
\providecommand \enquote  [1]{``#1''}%
\providecommand \bibnamefont  [1]{#1}%
\providecommand \bibfnamefont [1]{#1}%
\providecommand \citenamefont [1]{#1}%
\providecommand \href@noop [0]{\@secondoftwo}%
\providecommand \href [0]{\begingroup \@sanitize@url \@href}%
\providecommand \@href[1]{\@@startlink{#1}\@@href}%
\providecommand \@@href[1]{\endgroup#1\@@endlink}%
\providecommand \@sanitize@url [0]{\catcode `\\12\catcode `\$12\catcode
  `\&12\catcode `\#12\catcode `\^12\catcode `\_12\catcode `\%12\relax}%
\providecommand \@@startlink[1]{}%
\providecommand \@@endlink[0]{}%
\providecommand \url  [0]{\begingroup\@sanitize@url \@url }%
\providecommand \@url [1]{\endgroup\@href {#1}{\urlprefix }}%
\providecommand \urlprefix  [0]{URL }%
\providecommand \Eprint [0]{\href }%
\providecommand \doibase [0]{http://dx.doi.org/}%
\providecommand \selectlanguage [0]{\@gobble}%
\providecommand \bibinfo  [0]{\@secondoftwo}%
\providecommand \bibfield  [0]{\@secondoftwo}%
\providecommand \translation [1]{[#1]}%
\providecommand \BibitemOpen [0]{}%
\providecommand \bibitemStop [0]{}%
\providecommand \bibitemNoStop [0]{.\EOS\space}%
\providecommand \EOS [0]{\spacefactor3000\relax}%
\providecommand \BibitemShut  [1]{\csname bibitem#1\endcsname}%
\let\auto@bib@innerbib\@empty
\bibitem [{\citenamefont {Aoki}\ \emph {et~al.}(2006)\citenamefont {Aoki},
  \citenamefont {Dayan}, \citenamefont {Wilcut}, \citenamefont {Bowen},
  \citenamefont {Parkins}, \citenamefont {Kippenberg}, \citenamefont {Vahala},\
  and\ \citenamefont {Kimble}}]{Aoki2006}%
  \BibitemOpen
  \bibfield  {author} {\bibinfo {author} {\bibfnamefont {T.}~\bibnamefont
  {Aoki}}, \bibinfo {author} {\bibfnamefont {B.}~\bibnamefont {Dayan}},
  \bibinfo {author} {\bibfnamefont {E.}~\bibnamefont {Wilcut}}, \bibinfo
  {author} {\bibfnamefont {W.~P.}\ \bibnamefont {Bowen}}, \bibinfo {author}
  {\bibfnamefont {A.~S.}\ \bibnamefont {Parkins}}, \bibinfo {author}
  {\bibfnamefont {T.~J.}\ \bibnamefont {Kippenberg}}, \bibinfo {author}
  {\bibfnamefont {K.~J.}\ \bibnamefont {Vahala}}, \ and\ \bibinfo {author}
  {\bibfnamefont {H.~J.}\ \bibnamefont {Kimble}},\ }\href {\doibase
  10.1038/nature05147} {\bibfield  {journal} {\bibinfo  {journal} {Nature}\
  }\textbf {\bibinfo {volume} {443}},\ \bibinfo {pages} {671} (\bibinfo {year}
  {2006})}\BibitemShut {NoStop}%
\bibitem [{\citenamefont {Kippenberg}\ and\ \citenamefont
  {Vahala}(2008)}]{Kippenberg2008}%
  \BibitemOpen
  \bibfield  {author} {\bibinfo {author} {\bibfnamefont {T.~J.}\ \bibnamefont
  {Kippenberg}}\ and\ \bibinfo {author} {\bibfnamefont {K.~J.}\ \bibnamefont
  {Vahala}},\ }\href {http://dx.doi.org/10.1126/science.1156032} {\bibfield
  {journal} {\bibinfo  {journal} {Science}\ }\textbf {\bibinfo {volume}
  {321}},\ \bibinfo {pages} {1172} (\bibinfo {year} {2008})}\BibitemShut
  {NoStop}%
\bibitem [{\citenamefont {Vollmer}\ and\ \citenamefont
  {Arnold}(2008)}]{Vollmer2008}%
  \BibitemOpen
  \bibfield  {author} {\bibinfo {author} {\bibfnamefont {F.}~\bibnamefont
  {Vollmer}}\ and\ \bibinfo {author} {\bibfnamefont {S.}~\bibnamefont
  {Arnold}},\ }\href {\doibase 10.1038/nmeth.1221} {\bibfield  {journal}
  {\bibinfo  {journal} {Nat. methods}\ }\textbf {\bibinfo {volume} {5}},\
  \bibinfo {pages} {591} (\bibinfo {year} {2008})}\BibitemShut {NoStop}%
\bibitem [{\citenamefont {Spillane}\ \emph {et~al.}(2002)\citenamefont
  {Spillane}, \citenamefont {Kippenberg},\ and\ \citenamefont
  {Vahala}}]{Spillane2002}%
  \BibitemOpen
  \bibfield  {author} {\bibinfo {author} {\bibfnamefont {S.~M.}\ \bibnamefont
  {Spillane}}, \bibinfo {author} {\bibfnamefont {T.~J.}\ \bibnamefont
  {Kippenberg}}, \ and\ \bibinfo {author} {\bibfnamefont {K.~J.}\ \bibnamefont
  {Vahala}},\ }\href {\doibase 10.1038/415621a} {\bibfield  {journal} {\bibinfo
   {journal} {Nature}\ }\textbf {\bibinfo {volume} {415}},\ \bibinfo {pages}
  {621} (\bibinfo {year} {2002})}\BibitemShut {NoStop}%
\bibitem [{\citenamefont {Bahl}\ \emph {et~al.}(2011)\citenamefont {Bahl},
  \citenamefont {Zehnpfennig}, \citenamefont {Tomes},\ and\ \citenamefont
  {Carmon}}]{Bahl2011}%
  \BibitemOpen
  \bibfield  {author} {\bibinfo {author} {\bibfnamefont {G.}~\bibnamefont
  {Bahl}}, \bibinfo {author} {\bibfnamefont {J.}~\bibnamefont {Zehnpfennig}},
  \bibinfo {author} {\bibfnamefont {M.}~\bibnamefont {Tomes}}, \ and\ \bibinfo
  {author} {\bibfnamefont {T.}~\bibnamefont {Carmon}},\ }\href {\doibase
  10.1038/ncomms1412} {\bibfield  {journal} {\bibinfo  {journal} {Nat.
  Communications}\ }\textbf {\bibinfo {volume} {2}},\ \bibinfo {pages} {403}
  (\bibinfo {year} {2011})}\BibitemShut {NoStop}%
\bibitem [{\citenamefont {P{\"{o}}llinger}\ and\ \citenamefont
  {Rauschenbeutel}(2010)}]{Pollinger2010}%
  \BibitemOpen
  \bibfield  {author} {\bibinfo {author} {\bibfnamefont {M.}~\bibnamefont
  {P{\"{o}}llinger}}\ and\ \bibinfo {author} {\bibfnamefont {A.}~\bibnamefont
  {Rauschenbeutel}},\ }\href {\doibase 10.1364/OE.18.017764} {\bibfield
  {journal} {\bibinfo  {journal} {Opt. Express}\ }\textbf {\bibinfo {volume}
  {18}},\ \bibinfo {pages} {17764} (\bibinfo {year} {2010})}\BibitemShut
  {NoStop}%
\bibitem [{\citenamefont {Kippenberg}\ \emph {et~al.}(2011)\citenamefont
  {Kippenberg}, \citenamefont {Holzwarth},\ and\ \citenamefont
  {Diddams}}]{Kippenberg2011}%
  \BibitemOpen
  \bibfield  {author} {\bibinfo {author} {\bibfnamefont {T.~J.}\ \bibnamefont
  {Kippenberg}}, \bibinfo {author} {\bibfnamefont {R.}~\bibnamefont
  {Holzwarth}}, \ and\ \bibinfo {author} {\bibfnamefont {S.~A.}\ \bibnamefont
  {Diddams}},\ }\href {\doibase 10.1126/science.1193968} {\bibfield  {journal}
  {\bibinfo  {journal} {Science}\ }\textbf {\bibinfo {volume}
  {332}},\ \bibinfo {pages} {555} (\bibinfo {year} {2011})}\BibitemShut
  {NoStop}%
\bibitem [{\citenamefont {Herr}\ \emph {et~al.}(2013)\citenamefont {Herr},
  \citenamefont {Brasch}, \citenamefont {Jost}, \citenamefont {Wang},
  \citenamefont {Kondratiev}, \citenamefont {Gorodetsky},\ and\ \citenamefont
  {Kippenberg}}]{Herr2013}%
  \BibitemOpen
  \bibfield  {author} {\bibinfo {author} {\bibfnamefont {T.}~\bibnamefont
  {Herr}}, \bibinfo {author} {\bibfnamefont {V.}~\bibnamefont {Brasch}},
  \bibinfo {author} {\bibfnamefont {J.~D.}\ \bibnamefont {Jost}}, \bibinfo
  {author} {\bibfnamefont {C.~Y.}\ \bibnamefont {Wang}}, \bibinfo {author}
  {\bibfnamefont {N.~M.}\ \bibnamefont {Kondratiev}}, \bibinfo {author}
  {\bibfnamefont {M.~L.}\ \bibnamefont {Gorodetsky}}, \ and\ \bibinfo {author}
  {\bibfnamefont {T.~J.}\ \bibnamefont {Kippenberg}},\ }\href {\doibase
  10.1038/nphoton.2013.343} {\bibfield  {journal} {\bibinfo  {journal} {Nat.
  Photonics}\ }\textbf {\bibinfo {volume} {8}}\ \bibinfo {pages} {145-152}(\bibinfo {year}
  {2013})}\BibitemShut {NoStop}%
\bibitem [{\citenamefont {Sharping}\ \emph {et~al.}(2002)\citenamefont
  {Sharping}, \citenamefont {Fiorentino}, \citenamefont {Kumar},\ and\
  \citenamefont {Windeler}}]{Sharping2002}%
  \BibitemOpen
  \bibfield  {author} {\bibinfo {author} {\bibfnamefont {J.~E.}\ \bibnamefont
  {Sharping}}, \bibinfo {author} {\bibfnamefont {M.}~\bibnamefont
  {Fiorentino}}, \bibinfo {author} {\bibfnamefont {P.}~\bibnamefont {Kumar}}, \
  and\ \bibinfo {author} {\bibfnamefont {R.~S.}\ \bibnamefont {Windeler}},\
  }\href {\doibase 10.1364/OL.27.001675} {\bibfield  {journal} {\bibinfo
  {journal} {Opt. Lett.}\ }\textbf {\bibinfo {volume} {27}},\ \bibinfo
  {pages} {1675} (\bibinfo {year} {2002})}\BibitemShut {NoStop}%
\bibitem [{\citenamefont {Del'Haye}\ \emph {et~al.}(2007)\citenamefont
  {Del'Haye}, \citenamefont {Schliesser}, \citenamefont {Arcizet},
  \citenamefont {Wilken}, \citenamefont {Holzwarth},\ and\ \citenamefont
  {Kippenberg}}]{DelHaye2007}%
  \BibitemOpen
  \bibfield  {author} {\bibinfo {author} {\bibfnamefont {P.}~\bibnamefont
  {Del'Haye}}, \bibinfo {author} {\bibfnamefont {A.}~\bibnamefont
  {Schliesser}}, \bibinfo {author} {\bibfnamefont {O.}~\bibnamefont {Arcizet}},
  \bibinfo {author} {\bibfnamefont {T.}~\bibnamefont {Wilken}}, \bibinfo
  {author} {\bibfnamefont {R.}~\bibnamefont {Holzwarth}}, \ and\ \bibinfo
  {author} {\bibfnamefont {T.~J.}\ \bibnamefont {Kippenberg}},\ }\href
  {\doibase 10.1038/nature06401} {\bibfield  {journal} {\bibinfo  {journal}
  {Nature}\ }\textbf {\bibinfo {volume} {450}},\ \bibinfo {pages} {1214}
  (\bibinfo {year} {2007})}\BibitemShut {NoStop}%
\bibitem [{\citenamefont {Ilchenko}\ \emph {et~al.}(2003)\citenamefont
  {Ilchenko}, \citenamefont {Savchenkov}, \citenamefont {Matsko},\ and\
  \citenamefont {Maleki}}]{Ilchenko2003}%
  \BibitemOpen
  \bibfield  {author} {\bibinfo {author} {\bibfnamefont {V.~S.}\ \bibnamefont
  {Ilchenko}}, \bibinfo {author} {\bibfnamefont {A.~A.}\ \bibnamefont
  {Savchenkov}}, \bibinfo {author} {\bibfnamefont {A.~B.}\ \bibnamefont
  {Matsko}}, \ and\ \bibinfo {author} {\bibfnamefont {L.}~\bibnamefont
  {Maleki}},\ }\href {\doibase 10.1364/JOSAA.20.000157} {\bibfield  {journal}
  {\bibinfo  {journal} {J. Opt. Soc. Am. A}\ }\textbf
  {\bibinfo {volume} {20}},\ \bibinfo {pages} {157} (\bibinfo {year}
  {2003})}\BibitemShut {NoStop}%
\bibitem [{\citenamefont {Kippenberg}\ \emph {et~al.}(2004)\citenamefont
  {Kippenberg}, \citenamefont {Spillane},\ and\ \citenamefont
  {Vahala}}]{Kippenberg2004}%
  \BibitemOpen
  \bibfield  {author} {\bibinfo {author} {\bibfnamefont {T.~J.}\ \bibnamefont
  {Kippenberg}}, \bibinfo {author} {\bibfnamefont {S.~M.}\ \bibnamefont
  {Spillane}}, \ and\ \bibinfo {author} {\bibfnamefont {K.~J.}\ \bibnamefont
  {Vahala}},\ }\href {\doibase 10.1103/PhysRevLett.93.083904} {\bibfield
  {journal} {\bibinfo  {journal} {Phys. Rev. Lett.}\ }\textbf {\bibinfo
  {volume} {93}},\ \bibinfo {pages} {083904} (\bibinfo {year}
  {2004})}\BibitemShut {NoStop}%
\bibitem [{\citenamefont {Agha}\ \emph {et~al.}(2007)\citenamefont {Agha},
  \citenamefont {Okawachi}, \citenamefont {Foster}, \citenamefont {Sharping},\
  and\ \citenamefont {Gaeta}}]{Agha2007}%
  \BibitemOpen
  \bibfield  {author} {\bibinfo {author} {\bibfnamefont {I.}~\bibnamefont
  {Agha}}, \bibinfo {author} {\bibfnamefont {Y.}~\bibnamefont {Okawachi}},
  \bibinfo {author} {\bibfnamefont {M.}~\bibnamefont {Foster}}, \bibinfo
  {author} {\bibfnamefont {J.}~\bibnamefont {Sharping}}, \ and\ \bibinfo
  {author} {\bibfnamefont {A.}~\bibnamefont {Gaeta}},\ }\href {\doibase
  10.1103/PhysRevA.76.043837} {\bibfield  {journal} {\bibinfo  {journal}
  {Phys. Rev. A}\ }\textbf {\bibinfo {volume} {76}},\ \bibinfo {pages}
  {043837} (\bibinfo {year} {2007})}\BibitemShut {NoStop}%
\bibitem [{\citenamefont {Li}\ \emph {et~al.}(2013)\citenamefont {Li},
  \citenamefont {Wu}, \citenamefont {Liu},\ and\ \citenamefont {Xu}}]{Li2013}%
  \BibitemOpen
  \bibfield  {author} {\bibinfo {author} {\bibfnamefont {M.}~\bibnamefont
  {Li}}, \bibinfo {author} {\bibfnamefont {X.}~\bibnamefont {Wu}}, \bibinfo
  {author} {\bibfnamefont {L.}~\bibnamefont {Liu}}, \ and\ \bibinfo {author}
  {\bibfnamefont {L.}~\bibnamefont {Xu}},\ }\href {\doibase
  10.1364/OE.21.016908} {\bibfield  {journal} {\bibinfo  {journal} {Opt.
  Express}\ }\textbf {\bibinfo {volume} {21}},\ \bibinfo {pages} {16908}
  (\bibinfo {year} {2013})}\BibitemShut {NoStop}%
\bibitem [{\citenamefont {Gorodetsky}\ and\ \citenamefont
  {Fomin}(2006)}]{Gorodetsky2006}%
  \BibitemOpen
  \bibfield  {author} {\bibinfo {author} {\bibfnamefont {M.}~\bibnamefont
  {Gorodetsky}}\ and\ \bibinfo {author} {\bibfnamefont {A.}~\bibnamefont
  {Fomin}},\ }\href {\doibase 10.1109/JSTQE.2005.862954} {\bibfield  {journal}
  {\bibinfo  {journal} {IEEE J. Sel. Top. Quantum
  Electron.}\ }\textbf {\bibinfo {volume} {12}},\ \bibinfo {pages} {33}
  (\bibinfo {year} {2006})}\BibitemShut {NoStop}%
\bibitem [{\citenamefont {Sumetsky}(2004)}]{Sumetsky2004}%
  \BibitemOpen
  \bibfield  {author} {\bibinfo {author} {\bibfnamefont {M.}~\bibnamefont
  {Sumetsky}},\ }\href {\doibase 10.1364/OL.29.000008} {\bibfield  {journal}
  {\bibinfo  {journal} {Opt. Lett.}\ }\textbf {\bibinfo {volume} {29}},\
  \bibinfo {pages} {8} (\bibinfo {year} {2004})}\BibitemShut {NoStop}%
\bibitem [{\citenamefont {P{\"{o}}llinger}\ \emph {et~al.}(2009)\citenamefont
  {P{\"{o}}llinger}, \citenamefont {O'Shea}, \citenamefont {Warken},\ and\
  \citenamefont {Rauschenbeutel}}]{Pollinger2009}%
  \BibitemOpen
  \bibfield  {author} {\bibinfo {author} {\bibfnamefont {M.}~\bibnamefont
  {P{\"{o}}llinger}}, \bibinfo {author} {\bibfnamefont {D.}~\bibnamefont
  {O'Shea}}, \bibinfo {author} {\bibfnamefont {F.}~\bibnamefont {Warken}}, \
  and\ \bibinfo {author} {\bibfnamefont {A.}~\bibnamefont {Rauschenbeutel}},\
  }\href {\doibase 10.1103/PhysRevLett.103.053901} {\bibfield  {journal}
  {\bibinfo  {journal} {Phys. Rev. Lett.}\ }\textbf {\bibinfo {volume}
  {103}},\ \bibinfo {pages} {053901} (\bibinfo {year} {2009})}\BibitemShut
  {NoStop}%
\bibitem [{\citenamefont {Ooka}\ \emph {et~al.}(2015)\citenamefont {Ooka},
  \citenamefont {Yang}, \citenamefont {Ward},\ and\ \citenamefont
  {Chormaic}}]{Ooka2015}%
  \BibitemOpen
  \bibfield  {author} {\bibinfo {author} {\bibfnamefont {Y.}~\bibnamefont
  {Ooka}}, \bibinfo {author} {\bibfnamefont {Y.}~\bibnamefont {Yang}}, \bibinfo
  {author} {\bibfnamefont {J.}~\bibnamefont {Ward}}, \ and\ \bibinfo {author}
  {\bibfnamefont {S.}~\bibnamefont {Nic Chormaic}},\ }\href {\doibase
  10.7567/APEX.8.092001} {\bibfield  {journal} {\bibinfo  {journal} {Appl.
  Phys. Express}\ }\textbf {\bibinfo {volume} {8}},\ \bibinfo {pages}
  {092001} (\bibinfo {year} {2015})}\BibitemShut {NoStop}%
\bibitem [{\citenamefont {Ward}\ \emph {et~al.}(2006)\citenamefont {Ward},
  \citenamefont {O'Shea}, \citenamefont {Shortt}, \citenamefont {Morrissey},
  \citenamefont {Deasy},\ and\ \citenamefont {{Nic Chormaic}}}]{Ward2006}%
  \BibitemOpen
  \bibfield  {author} {\bibinfo {author} {\bibfnamefont {J.~M.}\ \bibnamefont
  {Ward}}, \bibinfo {author} {\bibfnamefont {D.}\ \bibnamefont {O'Shea}},
  \bibinfo {author} {\bibfnamefont {B.~J.}\ \bibnamefont {Shortt}}, \bibinfo
  {author} {\bibfnamefont {M.~J.}\ \bibnamefont {Morrissey}}, \bibinfo {author}
  {\bibfnamefont {K.}~\bibnamefont {Deasy}}, \ and\ \bibinfo {author}
  {\bibfnamefont {S.}\ \bibnamefont {{Nic Chormaic}}},\ }\href {\doibase
  10.1063/1.2239033} {\bibfield  {journal} {\bibinfo  {journal} {Rev. of
  Sci. Instr.}\ }\textbf {\bibinfo {volume} {77}},\ \bibinfo {pages}
  {083105} (\bibinfo {year} {2006})}\BibitemShut {NoStop}%
\bibitem [{\citenamefont {Farnesi}\ \emph {et~al.}(2015)\citenamefont
  {Farnesi}, \citenamefont {Barucci}, \citenamefont {Righini}, \citenamefont
  {Conti},\ and\ \citenamefont {Soria}}]{Farnesi2015}%
  \BibitemOpen
  \bibfield  {author} {\bibinfo {author} {\bibfnamefont {D.}~\bibnamefont
  {Farnesi}}, \bibinfo {author} {\bibfnamefont {A.}~\bibnamefont {Barucci}},
  \bibinfo {author} {\bibfnamefont {G.~C.}\ \bibnamefont {Righini}}, \bibinfo
  {author} {\bibfnamefont {G.~N.}\ \bibnamefont {Conti}}, \ and\ \bibinfo
  {author} {\bibfnamefont {S.}~\bibnamefont {Soria}},\ }\href {\doibase
  10.1364/OL.40.004508} {\bibfield  {journal} {\bibinfo  {journal} {Opt.
  Lett.}\ }\textbf {\bibinfo {volume} {40}},\ \bibinfo {pages} {4508}
  (\bibinfo {year} {2015})}\BibitemShut {NoStop}%
\bibitem [{\citenamefont {Min}\ \emph {et~al.}(2005)\citenamefont {Min},
  \citenamefont {Yang},\ and\ \citenamefont {Vahala}}]{Min2005}%
  \BibitemOpen
  \bibfield  {author} {\bibinfo {author} {\bibfnamefont {B.}~\bibnamefont
  {Min}}, \bibinfo {author} {\bibfnamefont {L.}~\bibnamefont {Yang}}, \ and\
  \bibinfo {author} {\bibfnamefont {K.}~\bibnamefont {Vahala}},\ }\href
  {\doibase 10.1063/1.2120921} {\bibfield  {journal} {\bibinfo  {journal}
  {Appl. Phys. Lett.}\ }\textbf {\bibinfo {volume} {87}},\ \bibinfo
  {pages} {181109} (\bibinfo {year} {2005})}\BibitemShut {NoStop}%
\bibitem [{\citenamefont {Sumetsky}\ and\ \citenamefont
  {Dulashko}(2012)}]{Sumetsky2012}%
  \BibitemOpen
  \bibfield  {author} {\bibinfo {author} {\bibfnamefont {M.}~\bibnamefont
  {Sumetsky}}\ and\ \bibinfo {author} {\bibfnamefont {Y.}~\bibnamefont
  {Dulashko}},\ }\href {\doibase 10.1364/OE.20.027896} {\bibfield  {journal}
  {\bibinfo  {journal} {Opt. Express}\ }\textbf {\bibinfo {volume} {20}},\
  \bibinfo {pages} {27896} (\bibinfo {year} {2012})}\BibitemShut {NoStop}%
\bibitem [{\citenamefont {Sumetsky}(2013)}]{Sumetsky2013}%
  \BibitemOpen
  \bibfield  {author} {\bibinfo {author} {\bibfnamefont {M.}~\bibnamefont
  {Sumetsky}},\ }\href {\doibase 10.1103/PhysRevLett.111.163901} {\bibfield
  {journal} {\bibinfo  {journal} {Phys. Rev. Lett.}\ }\textbf {\bibinfo
  {volume} {111}},\ \bibinfo {pages} {163901} (\bibinfo {year}
  {2013})}\BibitemShut {NoStop}%
\end{thebibliography}
%

\end{document}